\documentclass[aps,prd,amsmath,twocolumn,notitlepage,showpacs,superscriptaddress,nofootinbib,usenatbib,10pt]{revtex4-1}
\def\be{\begin{equation}}
\def\ee{\end{equation}}
\def\bea{\begin{eqnarray}}
\def\eea{\end{eqnarray}}

\usepackage{graphicx}% Include figure files 
\usepackage{dcolumn}% Align table columns on decimal point 
\usepackage{bm}% bold math 
\usepackage{epsfig} 
\usepackage{amsfonts}
\usepackage{amsmath}
\usepackage{amssymb}
\usepackage[usenames]{color}
\usepackage[colorlinks=true]{hyperref}

\hypersetup{colorlinks,citecolor=blue,linkcolor=blue,urlcolor=blue}
\hypersetup{final=true}

\begin{document}

\title{{The Hubble constant tension with next-generation galaxy surveys}}

\author{Carlos A. P. Bengaly}
\email{carlosap87@gmail.com}
\affiliation{D\'epartement de Physique Th\'eorique, Universit\'e de Gen\`eve, 24 quai Ernest Ansermet, 1211 Gen\'eve 4, Switzerland}
\affiliation{Department of Physics \& Astronomy, University of the Western Cape, Bellville 7535, South Africa}

\author{Chris Clarkson}
\email{chris.clarkson@qmul.ac.uk}
\affiliation{School of Physics \& Astronomy, Queen Mary, University of London, United Kingdom}
\affiliation{Department of Mathematics \& Applied Mathematics, University of Cape Town, Rondebosch 7701, South Africa}
\affiliation{Department of Physics \& Astronomy, University of the Western Cape, Bellville 7535, South Africa}

\author{Roy Maartens}
\email{roy.maartens@gmail.com}
\affiliation{Department of Physics \& Astronomy, University of the Western Cape, Bellville 7535, South Africa}
\affiliation{Institute of Cosmology \& Gravitation, University of Portsmouth, Portsmouth PO1 3FX, United Kingdom}

\date{\today}

\begin{abstract}
The rate at which the universe is expanding today is a fundamental parameter in cosmology which governs our understanding of structure formation and dark energy.
However, current measurements of the Hubble constant, $H_0$, show a significant tension ($\sim 4-6\sigma$) between early- and late-Universe observations.
There are ongoing efforts to check the diverse observational results and also to investigate possible theoretical ways to resolve the tension~-- which could point to radical extensions of the standard model. Here we demonstrate the potential of next-generation spectroscopic galaxy surveys to shed light on the Hubble constant tension.  
Surveys such as those with Euclid and the Square Kilometre Array (SKA) are expected to reach sub-percent precision on Baryon Acoustic Oscillation (BAO)  measurements of the Hubble parameter, with a combined redshift coverage of $0.1<z<3$. This wide redshift range, together with the high precision and low level of systematics in BAO measurements, mean that these surveys will provide independent and tight constraints on $H(z)$. These $H(z)$ measurements can be  extrapolated to $z = 0$ to provide constraints on $H_0$ using a non-parametric regression. To this end we deploy Gaussian processes and we find that Euclid-like surveys can reach $\sim$3\% precision  on $H_0$, with SKA-like intensity mapping surveys reaching $\sim$2\%. When we combine the low-redshift SKA-like Band 2 survey with either its high-redshift Band 1 counterpart, or with the non-overlapping Euclid-like survey, the precision is predicted to be close to 1\% with 40 $H(z)$ data points.  This would be sufficient to rule out the current early- or late-Universe measurements at a $\sim$5$\sigma$ level. 
\end{abstract}

\maketitle

%--------------------------------------------------------------------------------------------------------------------------
\section{Introduction}\label{sec:intro}

The Hubble constant $H_0 = 100h\, \mathrm{km \, s}^{-1} \, \mathrm{Mpc}^{-1}$ is a fundamental cosmological parameter requiring precise measurement.
However, there is a significant tension between   
the {\em Planck} measurement from cosmic microwave background (CMB) anisotropies, assuming a concordance model~\cite{planck18} (see also~\cite{Ade:2015xua}),
\begin{equation}\label{eq:h0_P18}
H_0^{\rm P18} = 67.36 \pm 0.54 ~~\mathrm{km \, s}^{-1} \, \mathrm{Mpc}^{-1} \,,
\end{equation}
and measurements using type Ia supernovae (SNIa)  calibrated with Cepheid distances~\cite{riess19} (see also~\cite{riess16, riess18a}),
\begin{equation}\label{eq:h0_R19}
H_0^{\rm R19} = 74.03 \pm 1.42 ~~ \mathrm{km \, s}^{-1} \, \mathrm{Mpc}^{-1} \,.
\end{equation}
  
Recent measurements using time delays from lensed quasars~\cite{wong19} obtained $H_0 = 73.3^{+1.7}_{-1.8} \, \mathrm{km \, s}^{-1} \, \mathrm{Mpc}^{-1}$,
while~\cite{Freedman:2019jwv} found $H_0 = 69.8 \pm 1.9\, \mathrm{km \, s}^{-1} \, \mathrm{Mpc}^{-1}$ using the tip of the red giant branch applied to SNIa, which is independent of the Cepheid distance scale - in contrast with the $H_0 = 72.4 \pm1.9\, \mathrm{km \, s}^{-1} \, \mathrm{Mpc}^{-1}$ measurement obtained by~\cite{yuan19}. Analysis of a compilation of these and other recent high- and low-redshift measurements shows~\cite{Verde:2019ivm} that the discrepancy between P18 and any three independent late-Universe measurements is between 4 and $6\sigma$.

Here our focus is not on ways to explain the tension  via possible observational systematics or theoretical modifications to the cosmological model, but on the potential of next-generation spectroscopic surveys to provide an independent way of ruling out the P18 or R19 measurement. We take R19 as the representative late-Universe measurement, but the method and results apply to other such recent measurements or combinations of them that are in tension with P18 at a level $\gtrsim 4\sigma$.

Next-generation  spectroscopic surveys will measure the redshift and angular extents of the baryon acoustic oscillation (BAO) feature, $\Delta z$ and $\Delta\theta$. 
The BAO radial and transverse  physical scales are $\Delta R_\|(z)={\Delta z(z)/ (1+z)H_\|(z)}$ and $\Delta R_\perp(z)=d_A(z)\Delta\theta(z)$,
where $z$ is the observed redshift, $H_\|$ is the radial rate of expansion of matter~\cite{Jimenez:2019cll}
and $d_A$ is  the angular diameter distance. These expressions apply in a general cosmology. 

In a perturbed Friedmann model, $H_\|=H$, {so that $\Delta R_\|(z)$ directly determines $H(z)$,} while $(1+z) d_A$  is an integral of $H^{-1}$, so that $\Delta R_\perp(z)$ also contains information about $H(z)$.  {If we have independent determinations of the radial and transverse scales, we can find $H(z)$ from BAO measurements.}
The radial and transverse BAO scales should be equal, {after accounting for projection and  Alcock-Paczynski effects~\cite{Lepori:2016rzi}:}
$\Delta R_\|=\Delta R_\perp\equiv R$.
{The physical BAO scale at decoupling $R_{\rm d}=R(z_{\rm d})$ is the sound horizon, which is estimated with high precision by {\em Planck}. This  estimate
is extremely
insensitive to physics at low redshifts, since $R_{\rm d}$ is determined by
the physical matter densities $\Omega_{\rm c}h^2, \Omega_{\rm b} h^2$, which are  fixed mainly by the relative heights of the CMB acoustic peaks. The estimate of $R_{\rm d}$ assumes the $\Lambda$CDM cosmology at high redshifts~\cite{Lemos:2018smw}.}

Spectroscopic surveys with Euclid~\cite{euclid18} and SKA1 (using 21cm intensity mapping)~\cite{ska18} are forecast to deliver errors on $H(z)$ that are sub-percent for $0.5\lesssim z\lesssim 2$ and $O(1)\%$ for lower and higher $z$  once systematics like foreground cleaning, in the case of intensity mapping, are properly taken into account~\cite{bull16,villaescusa-navarro16}. We take the forecast errors  on $H(z)$, over the redshift ranges of Euclid-like and SKA-like surveys,  from~\cite{ska18} (see the left panel of their Figure 10). Then we use a non-parametric Gaussian process  to estimate $H_0$ from a regression analysis on the mock $H(z)$ data points, assuming the standard flat $\Lambda$CDM model  (see also~\cite{busti14, wang17b, gomezvalent18}). The regression produces errors on $H_0$, which we compare to the errors from P18 and R19  in \eqref{eq:h0_P18} and \eqref{eq:h0_R19}.

In summary, we aim to answer the following questions: 

How precise are the $H_0$ estimates that Euclid- and SKA-like surveys can obtain?
Can these surveys rule out P18 or R19?

%--------------------------------------------------------------------------------------------------------------------------
\section{Data Analysis}\label{sec:DA} 

%--------------------------------------------------------------------------------------------------------------------------
\begin{figure}[!t]
\centering
\includegraphics[width=0.45\textwidth]{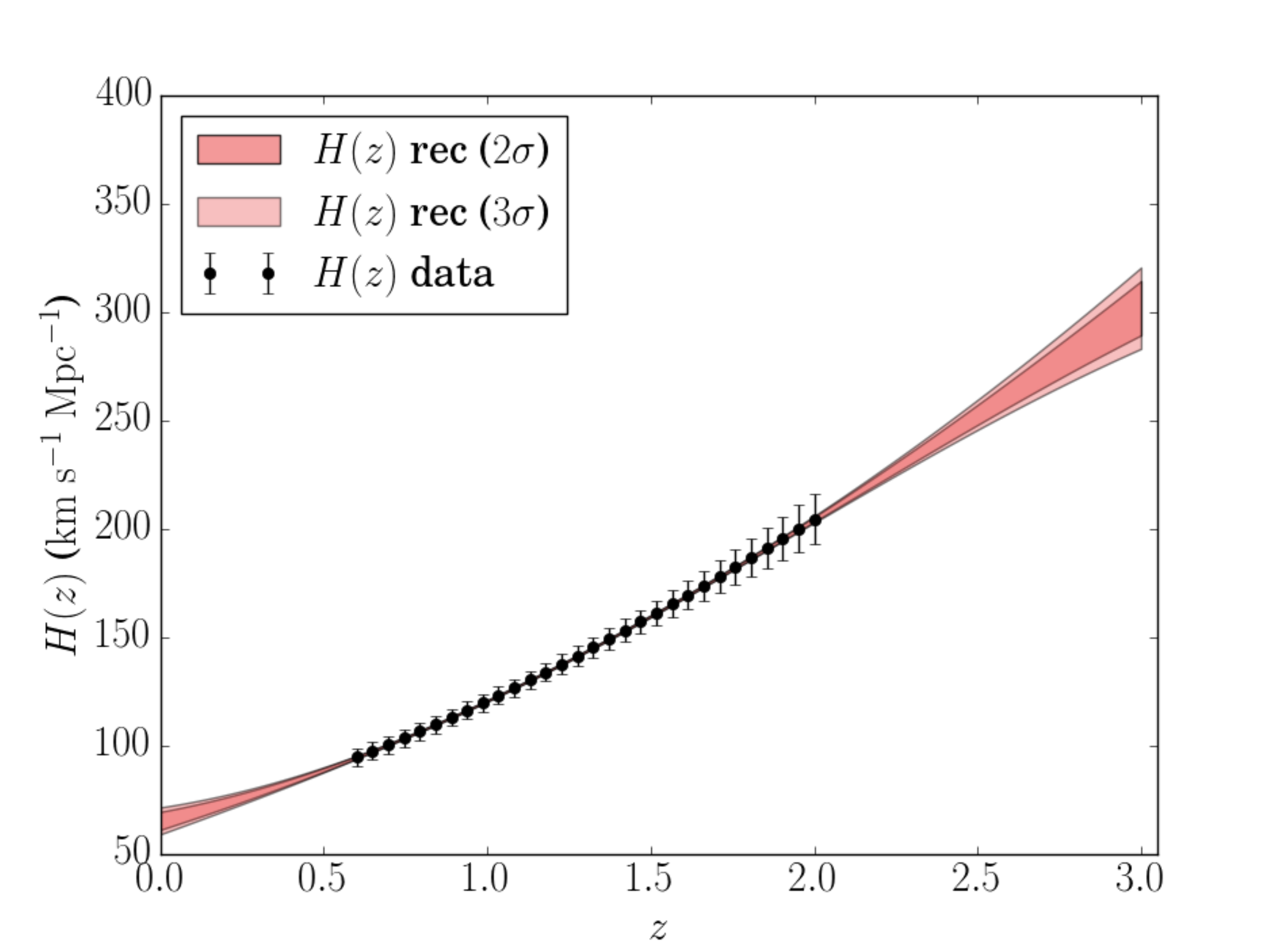}
\includegraphics[width=0.45\textwidth]{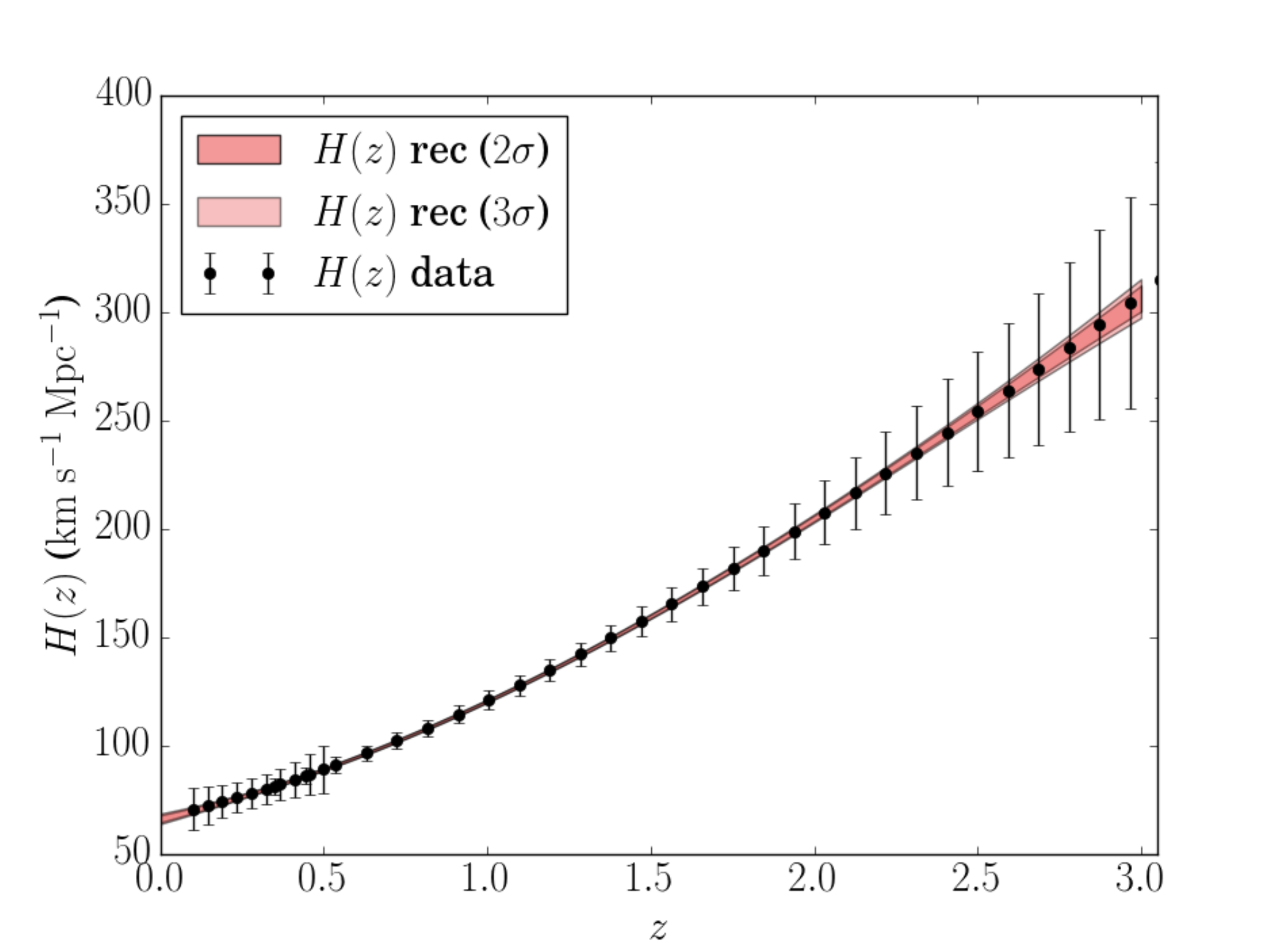}
\caption{Gaussian-process reconstructed $H(z)$ for Euclid-like survey \eqref{euc} (top) and SKA-like B1+B2 {(combined)} surveys \eqref{ska} (bottom), showing $2\sigma$ (darker) and $3\sigma$ (lighter) reconstruction uncertainty. We use $N=30$ (top) and $N_1=30, N_2=10$ (bottom), and a fiducial $H_0$ value assuming P18. Error bars ($1\sigma$) on data  points are increased by a factor 10 (top) and 6 (bottom) for visibility. } 
\label{fig:hz_euclid}
\end{figure}
%--------------------------------------------------------------------------------------------------------------------------

A Gaussian process is a distribution over functions, rather than over variables as in the case of a Gaussian distribution. This allows us to reconstruct a function from data points without assuming a parametrization. We use the {\sc GaPP} (Gaussian Processes in Python) code~\cite{seikel12} (see also~\cite{shafieloo12}) in order to reconstruct $H(z)$ from data. (For other applications of {\sc GaPP} in cosmology, see e.g.~\cite{yahya13, gonzalez16, pinho18, vonmarttens19}.)  

We simulate $H(z)$ data assuming the fiducial model, 
\begin{equation}\label{eq:hz}
H^{\rm fid}(z) = H^{\rm fid}_0\big[\Omega_{\rm m}(1+z)^3 + (1-\Omega_{\rm m})\big]^{1/2},
\end{equation}
where $H^{\rm fid}_0$ is chosen as either the P18 or R19 best-fit in \eqref{eq:h0_P18} or \eqref{eq:h0_R19}. We fix the matter density to the P18 best-fit (TT, TE, EE+lowE+lensing): 
\begin{equation}\label{eq:omega_m}
\Omega_{\rm m} = 0.3166 \pm 0.0084\,.
\end{equation}

The tension between two $H_0$ measurements is defined following~\cite{camarena18, bengaly19a} as
\begin{equation}\label{eq:th0} 
T ={\Delta h \over \sigma} = 
\frac{\big|h_1 - h_2\big|}{\left[\sigma(h_1)^2 + \sigma(h_2)^2\right]^{1/2}}\,,
\end{equation}
which is valid for an one-dimension Gaussian distribution.
The current tension between the measured values in \eqref{eq:h0_P18} and \eqref{eq:h0_R19} is $T_h=4.4$, corresponding to $\Delta h=4.4\sigma$.

When we apply \eqref{eq:th0} to the reconstructed {$h(z)$ from mock data}, with fiducials  given by $h^{\rm P18}$ and $h^{\rm R19}$, the uncertainties do not depend on the fiducial, and so they are the same. Then \eqref{eq:th0} becomes
\be\label{trec}
T_{\rm rec} = {\Delta h_{\rm rec} \over \sigma_{\rm rec}}  =
\frac{\big|h^{\rm R19}_{\rm rec} - h^{\rm P18}_{\rm rec}\big|}{\sigma_{\rm rec}}\,,
\ee
where  $h^{\rm P18}_{\rm rec}$ and $h^{\rm R19}_{\rm rec}$ are the $h$ measurements reconstructed from the mock data, each with the same uncertainty $\sigma_{\rm rec}$.

For the surveys, we use the redshift ranges given in~\cite{euclid18,ska18}, and we assume a range of values for $N$, the number of $H(z)$ data points, as follows.

\underline{Euclid-like galaxy survey}:  
\be \label{euc}
0.6<z<2.0\,, \quad
N=10, 15, 20, 25, 30\,.
\ee

\underline{SKA-like intensity mapping  survey}:
\bea
\mbox{Band 1:~~}   && 0.35<z<3.06\,, ~ N=10, 15, 20, 25, 30\,, \notag\\
\mbox{Band 2:~~}   &&  0.1~<z~<0.5~\,, ~ N=5, 8, 10\,, \notag\\ 
\mbox{Band 1+2:~~}   && N_1=10, 15, 20, 25, 30~\mbox{and}~  N_2=10 \,, \label{ska}
\eea
{where Band 1+2 delivers the combined constraining power of Band 2 with 10 data points and Band 1 with $N_1$ points.}

The $H(z)$ measurement uncertainties are taken from the interpolated curves in Figure 10 (left) of~\cite{ska18}. 
%--------------------------------------------------------------------------------------------------------------------------

\section{Results}\label{sec:results}

\subsection{Measurements of the Hubble Constant} 

The Gaussian-process  reconstructed $H(z)$  for Euclid-like and SKA-like  B1+2 surveys  is shown in Figure~\ref{fig:hz_euclid} by  the 2 and $3\sigma$ regions of uncertainty on the reconstruction.  The data points and their forecast $1\sigma$ error bars are also shown -- where the errors are increased by 10 (Euclid-like) and {6} (SKA-like) to enhance visibility. We show $N=30$ for Euclid-like, and  $N=30$ for SKA-like B1, with a fixed $N=10$ for SKA-like B2. In these Figures, $H_0^{\rm P18}$ is the fiducial;  using  $H_0^{\rm R19}$ only shifts the reconstructed $H_0$ upward, but has no effect on its uncertainty.

%=================================================================
\begin{table}[!h]
\begin{center}
\begin{tabular}{ccccc}
\hline
\hline
Survey &  ~~$N$~~ & ~~$\sigma_{H_0}/H_0~ (\%)$~~  & ~~$T_{\rm rec}$~~ 
\\
\hline
\hline
	     & 10 & $3.993$ & $1.669$ \\
             & 15 & $3.705$ & $1.798$ \\
Euclid-like  & 20 & $3.482$ & $1.914$ \\
	     & 25 & $3.303$ & $2.017$ \\
	     & 30 & $3.155$ & $2.112$ \\
\hline
	     & 10 & $2.482$ & $2.684$ \\
             & 15 & $2.321$ & $2.871$ \\
SKA-like B1    & 20 & $2.178$ & $3.059$ \\
	     & 25 & $2.057$ & $3.239$ \\
	     & 30 & $1.954$ & $3.410$ \\
\hline
	     & 5 & $2.498$ & $2.664$  \\
SKA-like B2    & 8 & $2.175$ & $3.063$  \\
             & 10 & $2.038$ & $3.270$ \\
\hline  
	     & 20 & $1.362$ & $4.895$ \\
             & 25 & $1.302$ & $5.118$ \\
SKA-like B1+B2  & 30 & $1.260$ & $5.288$ \\
{(combined)}	     & 35 & $1.227$ & $5.432$ \\
	     & 40 & $1.198$ & $5.561$ \\
\hline  
P18          & - & $0.802$ & $4.390$ \\ 
R19          & - & $1.981$ & $4.390$ 
\\ 
\hline
\hline
\end{tabular}
\end{center}
\caption{Uncertainty of reconstructed $H_0$ and resulting $H_0$ tension \eqref{trec}, for simulated data from surveys \eqref{euc} and \eqref{ska}.}
\label{tab:tab_perc_th0} 
\end{table}
%=================================================================

%--------------------------------------------------------------------------------------------------------------------------
\begin{figure}%[!t]
\centering
\includegraphics[width=0.4\textwidth]{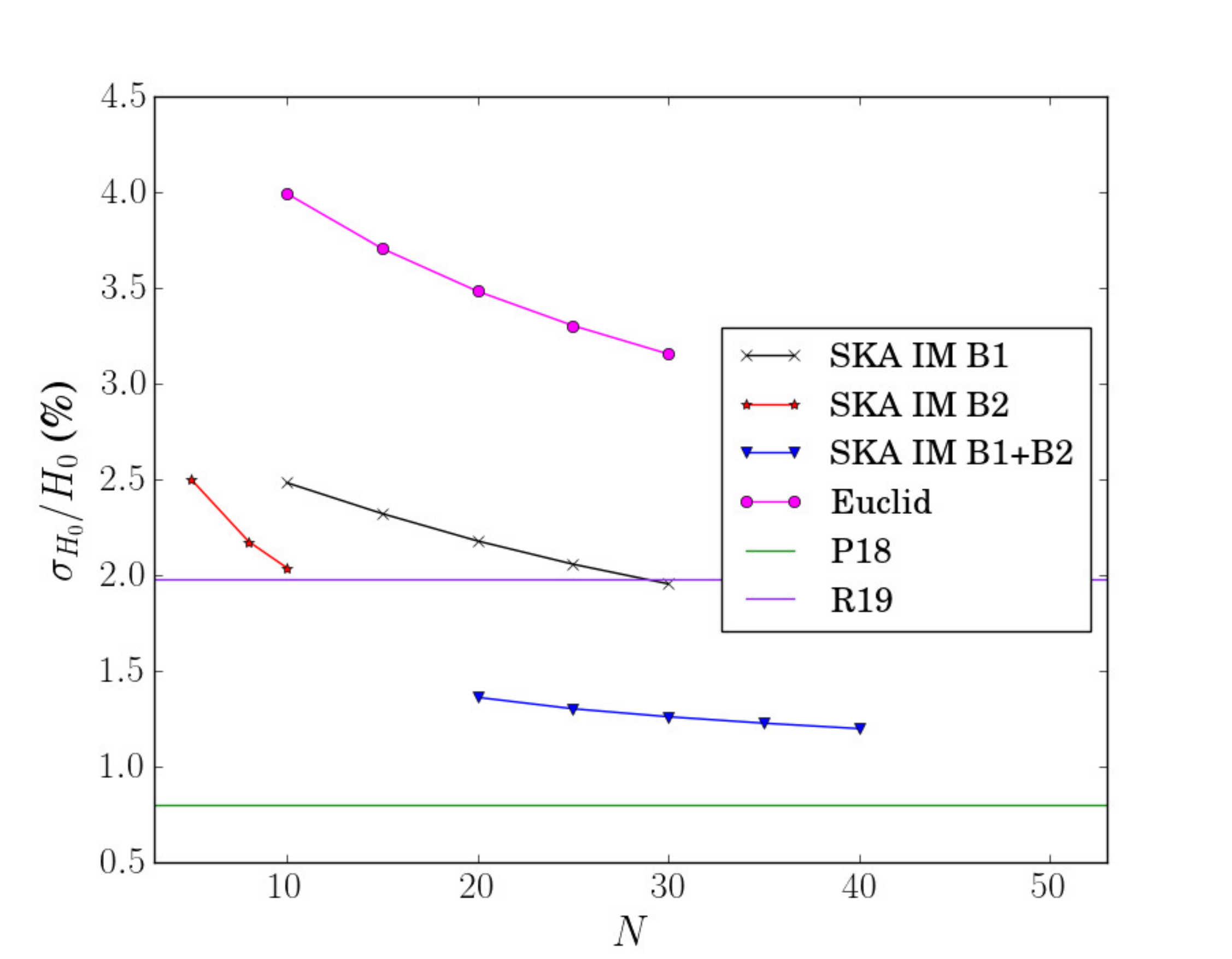}\\
\includegraphics[width=0.4\textwidth]{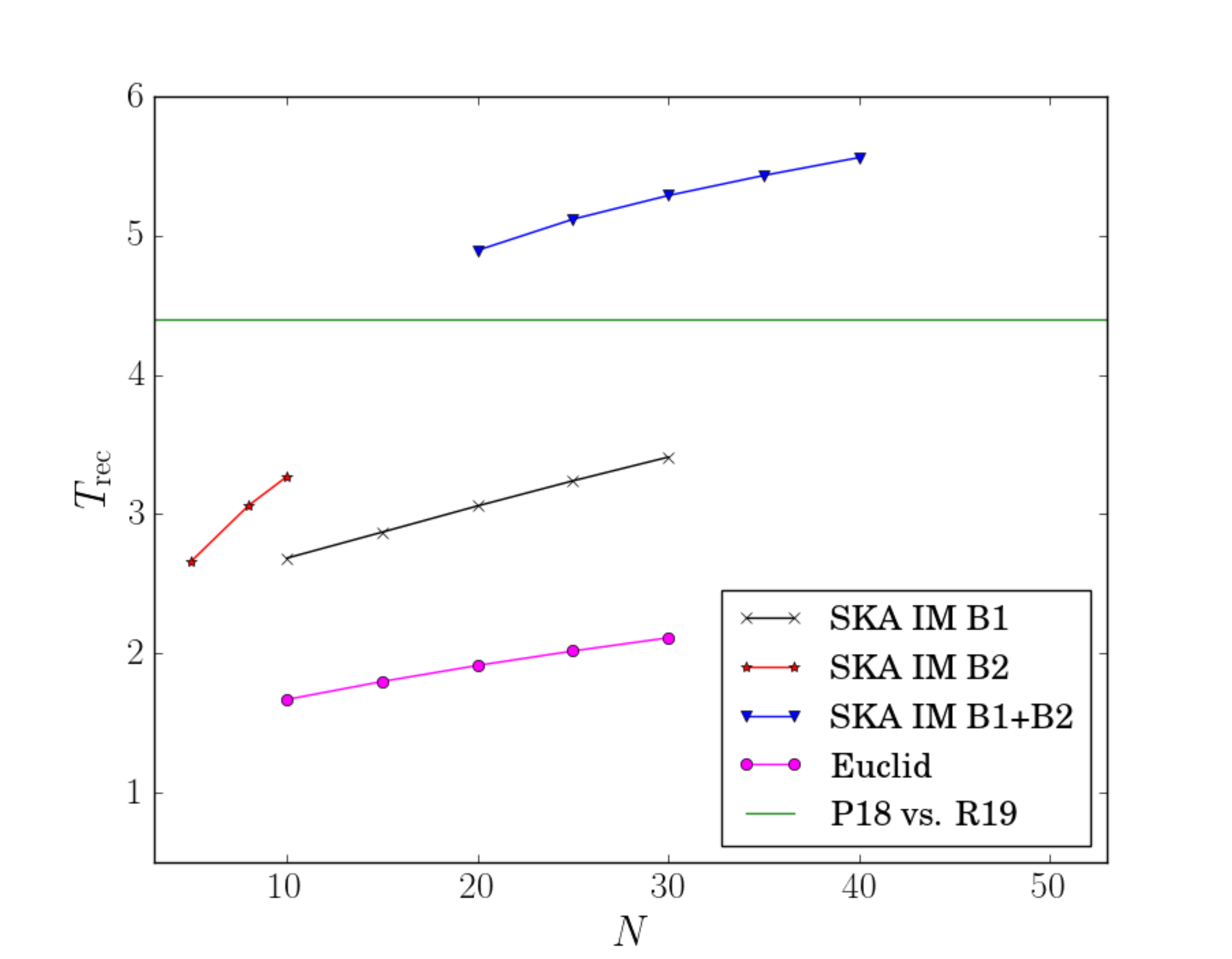}
\caption{{\it Top:}   Relative uncertainty on GP-reconstructed $H_0$ for Euclid- and SKA-like surveys with different  numbers of data points. Horizontal lines show P18 and R19 measurement uncertainties. {\it Bottom:} Corresponding tension \eqref{trec} between reconstructions with R19 and P18 fiducials. Horizontal line gives the tension between R19 and P18.} 
\label{fig:perc_th0}
\end{figure}
%--------------------------------------------------------------------------------------------------------------------------

%--------------------------------------------------------------------------------------------------------------------------
\begin{figure}%[!t]
\centering
\includegraphics[width=0.9\columnwidth]{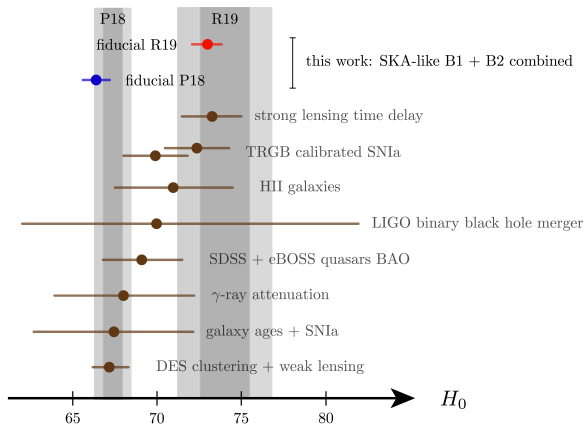}
\caption{Compilation of $H_0$ measurements, with $1\sigma$ error bars, shown against $1\sigma$ (darker) and $2\sigma$ (lighter) error bands for P18 (left) and R19 (right). From bottom to top:  DES clustering + weak lensing~\cite{abbott18};  
 galaxy ages + SNIa~\cite{gomezvalent18};
 $\gamma$-ray attenuation~\cite{dominguez19}; 
 SDSS + eBOSS quasars BAO (direct estimate of $H_0$)~\cite{wang17b};
 LIGO binary black hole merger GW170817~\cite{abbott17}; 
  HII galaxies~\cite{fernandezarenas18}; 
   TRGB calibrated SNIa{~\cite{Freedman:2019jwv, yuan19}};
 strong lensing time delay~\cite{wong19}.      
 Our GP-reconstructed  estimates for SKA-like B1+B2 {combined} are: (fiducial P18, in blue) and  (fiducial R19, in red), {where the dots indicate the reconstructed $H_0^{\rm P18}$ and  $H_0^{\rm R19}$}.} 
\label{fig:h0_measurements}
\end{figure}
%--------------------------------------------------------------------------------------------------------------------------

The reconstructed $H_0$ and its uncertainty follow from the intersection of the reconstruction  region with $z=0$ in Figure~\ref{fig:hz_euclid}, in the case of  $H_0^{\rm P18}$  as fiducial (and similarly for $H_0^{\rm R19}$ as fiducial). 

The results for the uncertainty and the tension are shown in Table~\ref{tab:tab_perc_th0} and  illustrated in Figure~\ref{fig:perc_th0}.  The individual SKA-like  B1 and B2 surveys perform better than the Euclid-like survey, given that  the former includes lower  and higher redshifts than the latter. 
With 10 low-$z$ (B2) data points and 30  high-$z$  (B1) points, SKA-like  surveys {in B1 and B2 separately} can provide $H_0$ measurements as precise as R19 ($\sim 2.0\%$ precision). Nonetheless, they are only able to discriminate between R19 and P18 at $\sim$3$\sigma$. 

On the other hand, the combination of B1 +  B2 can push  $\sigma_{H_0}/H_0$ down to $1.2\%$, which is close  to the P18 precision. This means that SKA-like B1+2 {combined} is predicted to be able to discriminate between P18 and R19 with $\sim$5$\sigma$ precision.

The results for SKA- and Euclid-like surveys are competitive with future standard siren measurements from gravitational wave events, whose forecasts predict a $H_0$ measurement with few percent precision~\cite{chen18, nair18, vitale18, mortlock18, shafieloo18, zhang19}. It is estimated that 50 binary neutron star standard sirens could resolve the P18--R19 tension~\cite{feeney19}, comparable to the 10 + 30 $H(z)$ measurements needed by SKA-like B1+2 {combined} surveys.   

Other model-independent methods are: $\gamma$-ray attenuation data~\cite{dominguez19}, giving a measurement with $\sigma_{H_0}/H_0=6.2\%$; Gaussian process regression on galaxy age determination of $H(z)$ together with SNIa data, giving~\cite{gomezvalent18} $\sigma_{H_0}/H_0=7.1\%$; HII galaxy data~\cite{fernandezarenas18}, giving $\sigma_{H_0}/H_0=4.9\%$. Methods that depend on assuming a cosmological model can deliver greater precision, but at the cost of losing model-independence. These include: using SDSS and eBOSS (quasars) BAO data, giving a direct H0 measurement (marginalising over $\Omega_m$)~\cite{wang17a} with $\sigma_{H_0}/H_0=3.4\%$; using DES clustering combined with weak lensing provides $\sigma_{H_0}/H_0=1.8\%$. Figure 3 displays these and other measurements, including our forecasts for SKA-like B1+B2 combined surveys with P18 and R19 fiducials.

\subsection{Robustness of results} 

The non-parametric Gaussian process regression that we use is not significantly sensitive to the cosmology assumed to perform the $H_0$ estimate. We verified this for dynamical dark energy extensions of the standard model, using $w$CDM and $(w_0,w_a)$CDM models. For example, the SKA-like B1+2 {combined} survey with 40 data points gives $\sigma_{H_0}/H_0=1.29\%$ with $(w_0,w_a) = (-1.1,-0.20)$ and $1.26\%$ with $(w_0,w_a) = (-0.9,+0.2)$. This is compatible with $\sigma_{H_0}/H_0=1.20\%$ obtained from the fiducial $\Lambda$CDM model (see Table~\ref{tab:tab_perc_th0}).  These results are consistent with the findings of~\cite{keeley19}, whose reconstructed cosmological parameters from GP are found to be unbiased with respect to the cosmological model assumed -- unlike parameter inference using methods like Monte Carlo Markov Chain.

We varied the cosmological parameters $\bm{p}=(\Omega_{\rm m},H_0)$ according to a Gaussian distribution $\mathcal{N}(\bm{p}, \sigma_{\bm{p}})$, where the parameters and their uncertainties are given by \eqref{eq:h0_P18}, \eqref{eq:h0_R19} and \eqref{eq:omega_m}. This test checks whether variations around the fiducial model within the limits imposed by state-of-the-art observations can bias our results. We find a negligible effect on the reconstructed  $H_0$ error, with an extra variation of only $\Delta(\sigma_{H_0}/H_0) \lesssim 0.1\%$ for SKA-like B1, which produces a change of only $\Delta T_{\rm rec} \sim 0.03\sigma_{\rm rec}$.  The results are qualitatively similar for the other surveys.  

We verified the robustness of our results with respect to changes  of the GP covariance function. By changing the squared exponential kernel that we used to the Mat\'ern(5/2), (7/2) and (9/2) kernels~\cite{seikel12,shafieloo12}, we obtained $\sigma_{H_0}/H_0=1.62\%,1.37\%,1.29\%$, for  SKA-like B1+B2 {combined}  with 25 data points. This is comparable to $1.30\%$ obtained in Table~\ref{tab:tab_perc_th0}. 

\begin{figure*}[!t]
\centering
\includegraphics[width=0.32\textwidth]{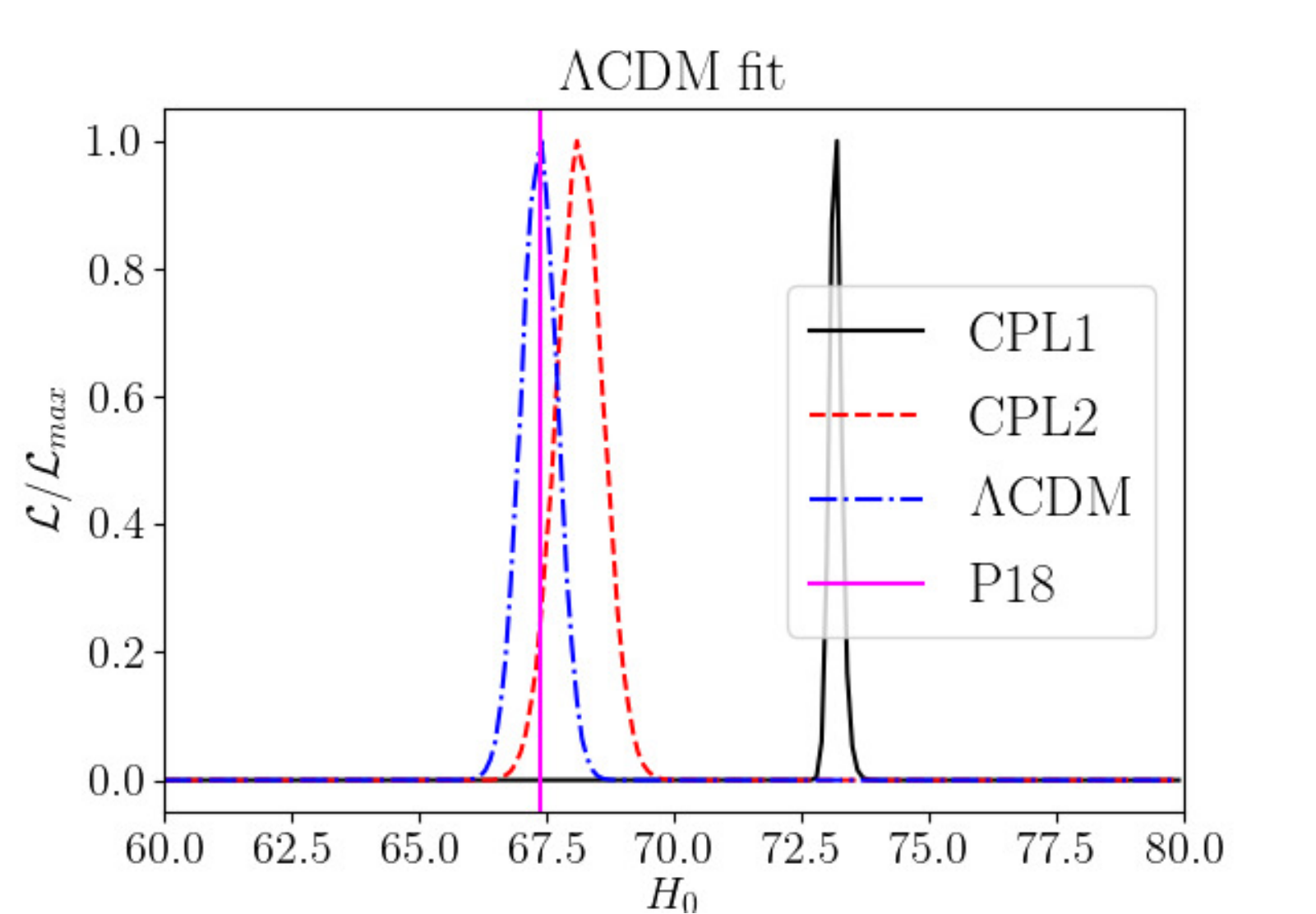}
\includegraphics[width=0.32\textwidth]{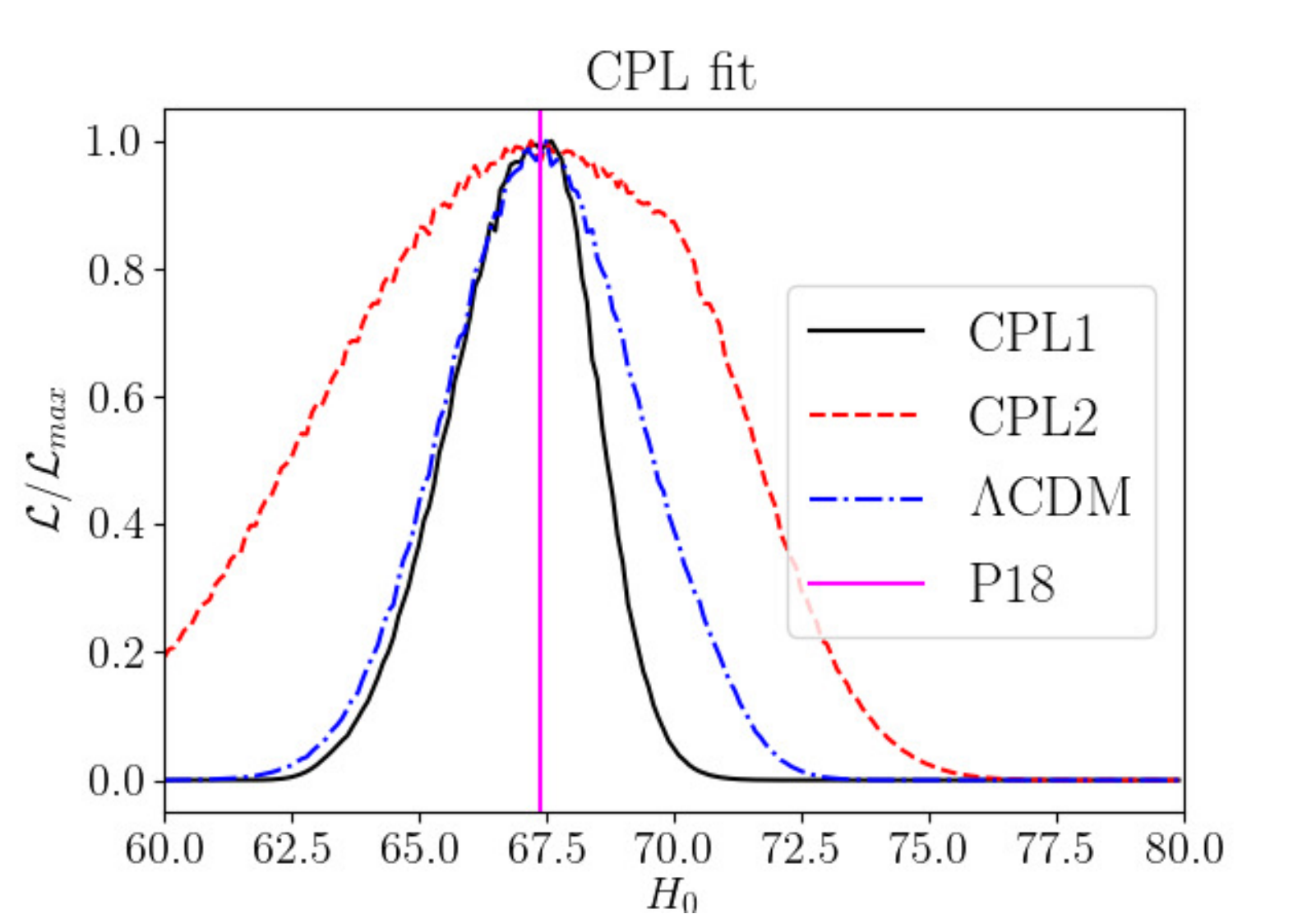}
\includegraphics[width=0.32\textwidth]{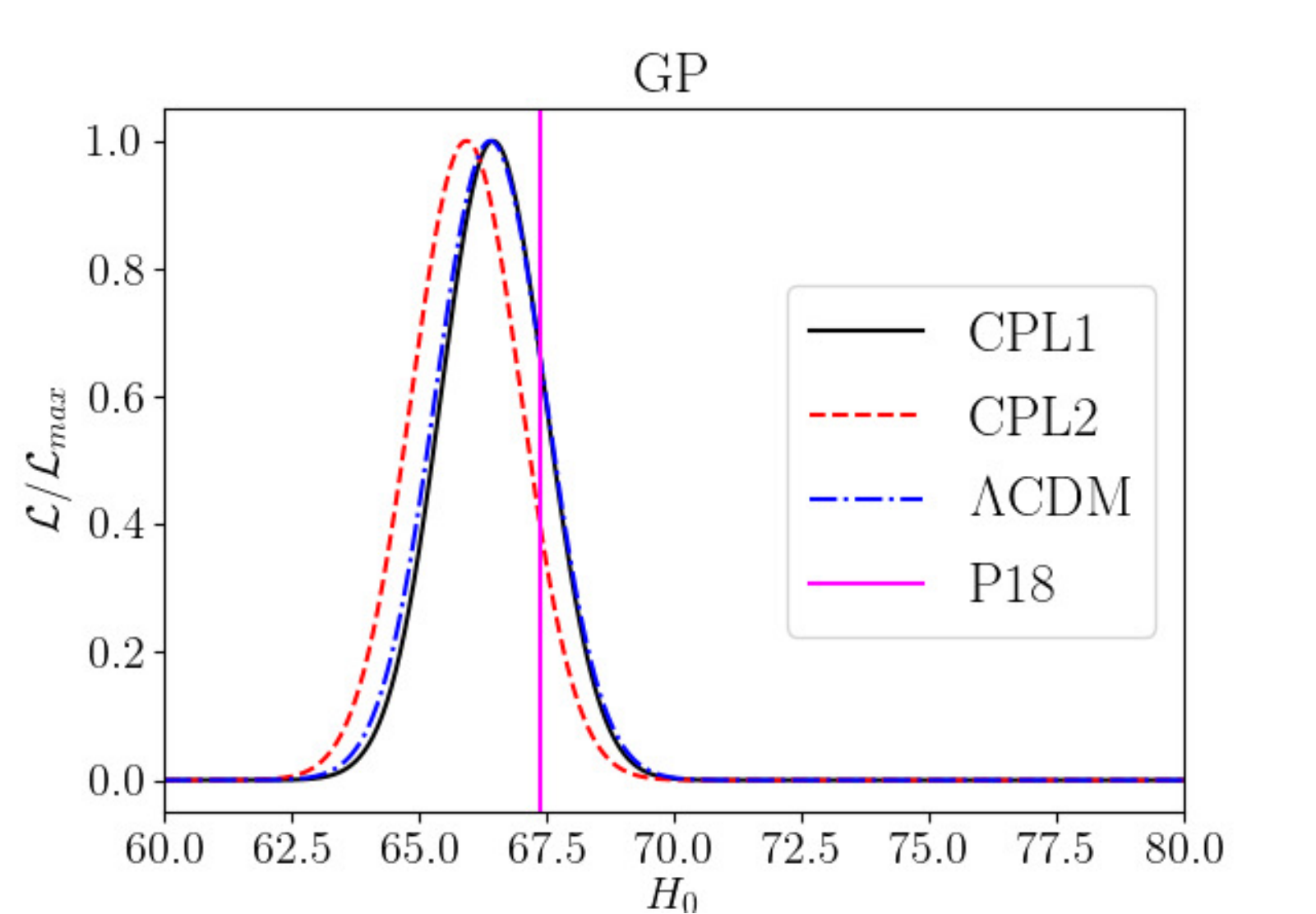}
\caption{Marginalised likelihood for the parameter $H_0$ by fitting a $H(z)$ simulation of SKA-like IM-B1+B2, 40 data points total, assuming three different cosmological models (see text) to $\Lambda$CDM (left) and CPL (middle). Right shows the recovered $H_0$ distribution using GP (with a Matern 5/2 covariance function).} 
\label{fig:likelihood_models}
\end{figure*}

As a final illustration of the effectiveness of our method, we show in Fig.~\ref{fig:likelihood_models} the recovered $H_0$ using standard parametric methods, assuming a $\Lambda$CDM and the CPL dark energy parameterization [$w(z) = w_0 + w_a (1-a)$], for simulations of a SKA-like IM B1+B2 surveys with 40 data points total. Assuming 3 different fiducial models ($\Lambda$CDM as the baseline model, while CPL1 has $w_0=-0.7,w_a=+0.4$ and CPL2 $w_0=-1.0,w_a=+0.4$), we see quite different results. For a $\Lambda$CDM parameterization, the errors are small ($\sigma H_0/H_0=0.56\%$), but can be significantly biased. Fitting a CPL model does recover the fiducial $H_0$ but at the expense of much larger errors ($\sigma_{H_0}/H_0=2.67\%$ for a $\Lambda$CDM simulation and $\sigma_{H_0}/H_0=5.95\%$ for CPL1, for instance), and fitting a ``wrong'' model (like fitting a $\Lambda$CDM model on a CPL simulation) yields biased results. The GP method recovers the correct value to within 1$\sigma$ irrespective of model, with the errors quoted above.

%---------------------------------------------------------------------------------------------------------------------------
\section{Conclusions}\label{sec:conc}
Next-generation spectroscopic surveys are expected to provide high precision BAO measurements, delivering  Hubble rate $H(z)$ data  over a wide range of $z$ with a low level of systematics. We estimated the potential precision from such measurements with Euclid-like and SKA-like surveys in estimating $H_0$ from $H(z)$, using non-parametric reconstruction and regression. We simulated $H(z)$ data sets following the expected specifications for both surveys, and carried out a Gaussian-process reconstruction of $H(z)$ from these data, allowing for regression down to $z=0$.  We checked the robustness of our results with respect to changes in the cosmological model and in the GP covariance functions. 

We found that SKA intensity mapping in Bands 1 and 2 {separately} can measure $H_0$ with $2.0-2.5\%$ precision, better than Euclid-like surveys with $3.2-4.0\%$ precision. 
Although these measurements are not able distinguish between the $H_0$ values from CMB and standard candles at higher than $3.5\sigma$, we found that the combination of SKA-like Band 1+2  can reach a precision of $1.2\% (1.4\%)$ with {20 (40)} total data points. This leads to a $4.9 (5.6)\sigma$ precision in distinguishing between the current P18 and R19 measurements.

The successful combination of low- and high-redshift data in SKA-like  surveys suggests an alternative: a combination of SKA-like Band 2 with Euclid-like surveys, which have no overlap between them. With $N_2=10$ Band 2 data points, and $N_{\rm E}$ Euclid data points, we find that the combined  constraining power is:
\be
{{\sigma_{H_0}\over H_0}(\mbox{\small SKA B2 + Euclid)}} = \begin{cases} 1.37\% &   N_{\rm E}=10\,,\\
1.32\% &   N_{\rm E}=20 \,,\\
1.29\% &  N_{\rm E}=30 \,.
\end{cases}
\ee
{This corresponds to a $4.9 (5.2)\sigma$ precision in distinguishing between the current P18 and R19 measurements with  {20 (40)} total data points. In other words, the combination of low-$z$ SKA- and high-$z$ Euclid-like surveys delivers precision that is almost as high as SKA-like B1+2.}

For comparison, we also computed the precision predicted  for other spectroscopic  surveys: the DESI galaxy survey~\cite{desi16}, a MeerKAT L-band intensity mapping survey~\cite{Santos:2017qgq, Santos2019}, and an SKA1 HI galaxy survey~\cite{ska18}. 
We find that:
\be
{\sigma_{H_0}\over H_0} \simeq \begin{cases} 10\% &  \mbox{DESI},\qquad~  0.6<z<2\,,~~~\quad N=30\,,\\
5\% &  \mbox{MeerKAT},~ 0.1<z<0.58\,, ~~ N=10 \,,\\
5\% &  \mbox{SKA GS}, \quad 0.1<z<0.5\,, ~~~\, N=10 \,.
\end{cases}
\ee

We conclude that Euclid-like galaxy and SKA-like intensity mapping surveys are forecast to provide the best $H_0$ estimates from GP regression of $H(z)$ measurements, allowing in the best cases (SKA-like B1+2, Euclid-like + SKA-like B2) for resolution of the tension between the $H_0$ measured from early- and late-Universe probes.\\

\emph{Acknowledgments} -- 
The authors acknowledge the anonymous referee for constructive criticism. CB and RM were supported by by the South African Radio Astronomy Observatory (SARAO) and the National Research Foundation (Grant No. 75415). CB also acknowledges support from the Swiss National Science Foundation at the late stage of this work. CC and RM were supported by the UK Science \& Technology Facilities Council Consolidated Grants ST/P000592/1 (CC) and ST/N000668/1 (RM). 
%--------------------------------------------------------------------------------------------------------------------------


\begin{thebibliography}{99}   

\bibitem{planck18} 
  N.~Aghanim {\it et al.} [Planck Collaboration],
   ``Planck 2018 results. VI. Cosmological parameters,''
  arXiv:1807.06209.

%\cite{Ade:2015xua}
\bibitem{Ade:2015xua} 
  P.~A.~R.~Ade {\it et al.} [Planck Collaboration],
  ``Planck 2015 results. XIII. Cosmological parameters,''
  Astron.\ Astrophys.\  {\bf 594}, A13 (2016)
   [arXiv:1502.01589].

\bibitem{riess19} 
  A.~G.~Riess, S.~Casertano, W.~Yuan, L.~M.~Macri and D.~Scolnic,
  ``Large Magellanic Cloud Cepheid Standards Provide a 1\% Foundation for the Determination of the Hubble Constant and Stronger Evidence for Physics beyond $\Lambda$CDM,''
  Astrophys.\ J.\  {\bf 876}, 85 (2019)
  [arXiv:1903.07603].  
  
\bibitem{riess18a} 
  A.~G.~Riess {\it et al.},
   ``Milky Way Cepheid Standards for Measuring Cosmic Distances and Application to Gaia DR2: Implications for the Hubble Constant,''
  Astrophys.\ J.\  {\bf 861},  126 (2018)
   [arXiv:1804.10655].
 
\bibitem{riess16} 
  A.~G.~Riess {\it et al.},
   ``A 2.4\% Determination of the Local Value of the Hubble Constant,''
  Astrophys.\ J.\  {\bf 826},  56 (2016)
  [arXiv:1604.01424].
  
\bibitem{wong19} 
  K.~C.~Wong {\it et al.},
   ``H0LiCOW XIII. A 2.4\% measurement of $H_{0}$ from lensed quasars: $5.3\sigma$ tension between early and late-Universe probes,''
  arXiv:1907.04869.
  
\bibitem{Freedman:2019jwv} 
  W.~L.~Freedman {\it et al.},
  ``The Carnegie-Chicago Hubble Program. VIII. An Independent Determination of the Hubble Constant Based on the Tip of the Red Giant Branch,''
  Astrophys.\ J.\  {\bf 881}, 1 (2019)
  [arXiv:1907.05922].

\bibitem{yuan19} 
  W.~Yuan, A.~G.~Riess, L.~M.~Macri, S.~Casertano and D.~Scolnic,
  ``Consistent Calibration of the Tip of the Red Giant Branch in the Large Magellanic Cloud on the Hubble Space Telescope Photometric System and Implications for the Determination of the Hubble Constant,''
  Astrophys.\ J.\  {\bf 886}, 61 (2019)
  [arXiv:1908.00993].
  
\bibitem{Verde:2019ivm} 
  L.~Verde, T.~Treu and A.~G.~Riess,
  ``Tensions between the Early and the Late Universe,''
  Nat. Astron. {\bf 3}, 891 (2019)
%   doi:10.1038/s41550-019-0902-0
  [arXiv:1907.10625].
  
\bibitem{Jimenez:2019cll} 
  R.~Jimenez, R.~Maartens, A.~R.~Khalifeh, R.~R.~Caldwell, A.~F.~Heavens and L.~Verde,
   ``Measuring the Homogeneity of the Universe Using Polarization Drift,''
  JCAP {\bf 1905}, 048 (2019)
%  doi:10.1088/1475-7516/2019/05/048
   [arXiv:1902.11298].

\bibitem{Lepori:2016rzi} 
  F.~Lepori, E.~Di Dio, M.~Viel, C.~Baccigalupi and R.~Durrer,
   ``The Alcock Paczynski test with Baryon Acoustic Oscillations: systematic effects for future surveys,''
  JCAP {\bf 1702}, 020 (2017)
 % doi:10.1088/1475-7516/2017/02/020
   [arXiv:1606.03114].
 
\bibitem{Lemos:2018smw} 
  P.~Lemos, E.~Lee, G.~Efstathiou and S.~Gratton,
  ``Model independent $H(z)$ reconstruction using the cosmic inverse distance ladder,''
  Mon.\ Not.\ Roy.\ Astron.\ Soc.\  {\bf 483},  4803 (2019)
%  doi:10.1093/mnras/sty3082
  [arXiv:1806.06781].
  
\bibitem{euclid18} 
  L.~Amendola {\it et al.},
   ``Cosmology and fundamental physics with the Euclid satellite,''
  Living Rev.\ Rel.\  {\bf 21},  2 (2018)
%   doi:10.1007/s41114-017-0010-3
   [arXiv:1606.00180].
  
\bibitem{ska18} 
  D.~J.~Bacon {\it et al.} [SKA Collaboration],
   ``Cosmology with Phase 1 of the Square Kilometre Array: Red Book 2018: Technical specifications and performance forecasts,''
   Publ. Astron. Soc. Austral. \textbf{37} (2020), e007 
   [arXiv:1811.02743].
      
\bibitem{bull16} 
  P.~Bull,
  ``Extending cosmological tests of General Relativity with the Square Kilometre Array,''
  Astrophys.\ J.\  {\bf 817}, no. 1, 26 (2016)
%   doi:10.3847/0004-637X/817/1/26
  [arXiv:1509.07562].
  
\bibitem{villaescusa-navarro16}
  F.~Villaescusa-Navarro, D.~Alonso and M.~Viel,
  ``Baryonic acoustic oscillations from 21 cm intensity mapping: the Square Kilometre Array case,''
  Mon.\ Not.\ Roy.\ Astron.\ Soc.\  {\bf 466} (2017) no.3,  2736
%   doi:10.1093/mnras/stw3224
  [arXiv:1609.00019].
  
\bibitem{busti14} 
  V.~C.~Busti, C.~Clarkson and M.~Seikel,
   ``Evidence for a Lower Value for $H_0$ from Cosmic Chronometers Data?,''
  Mon.\ Not.\ Roy.\ Astron.\ Soc.\  {\bf 441}, 11 (2014)
%   doi:10.1093/mnrasl/slu035
    [arXiv:1402.5429].

\bibitem{wang17b} 
 D.~Wang and X.~H.~Meng,
  ``Model-independent determination of $H_{0}$ using the latest cosmic chronometer data,''
 Sci.\ China Phys.\ Mech.\ Astron.\  {\bf 60},  110411 (2017)
%   doi:10.1007/s11433-017-9079-1
   [arXiv:1610.01202].
  
\bibitem{gomezvalent18} 
  A.~G\'omez-Valent and L.~Amendola,
   ``$H_0$ from cosmic chronometers and Type Ia supernovae, with Gaussian Processes and the novel Weighted Polynomial Regression method,''
  JCAP {\bf 1804}, 051 (2018)
%   doi:10.1088/1475-7516/2018/04/051
    [arXiv:1802.01505].
   
\bibitem{seikel12} 
  M.~Seikel, C.~Clarkson and M.~Smith,
   ``Reconstruction of dark energy and expansion dynamics using Gaussian processes,''
  JCAP {\bf 1206}, 036 (2012)
%   doi:10.1088/1475-7516/2012/06/036
    [arXiv:1204.2832].
    GaPP available at \url{http://www.acgc.uct.ac.za/~seikel/GAPP/index.html}
  
\bibitem{shafieloo12} 
  A.~Shafieloo, A.~G.~Kim and E.~V.~Linder,
  ``Gaussian Process Cosmography,''
  Phys.\ Rev.\ D {\bf 85}, 123530 (2012)
%   doi:10.1103/PhysRevD.85.123530
    [arXiv:1204.2272].
  
\bibitem{yahya13}
  S.~Yahya, M.~Seikel, C.~Clarkson, R.~Maartens and M.~Smith,
   ``Null tests of the cosmological constant using supernovae,''
  Phys.\ Rev.\ D {\bf 89},  023503 (2014)
%   doi:10.1103/PhysRevD.89.023503
  [arXiv:1308.4099].
    
\bibitem{gonzalez16} 
  J.~E.~Gonzalez, J.~S.~Alcaniz and J.~C.~Carvalho,
   ``Non-parametric reconstruction of cosmological matter perturbations,''
  JCAP {\bf 1604}, 016 (2016)
%   doi:10.1088/1475-7516/2016/04/016
    [arXiv:1602.01015].
  
\bibitem{pinho18} 
  A.~M.~Pinho, S.~Casas and L.~Amendola,
   ``Model-independent reconstruction of the linear anisotropic stress $\eta$,''
   JCAP {\bf 1811}, 027 (2018)
%   doi:10.1088/1475-7516/2018/11/027
  [arXiv:1805.00027].

\bibitem{vonmarttens19} 
  R.~von Marttens, V.~Marra, L.~Casarini, J.~E.~Gonzalez and J.~Alcaniz,
  ``Null test for interactions in the dark sector,''
  Phys.\ Rev.\ D {\bf 99},  043521 (2019)
%   doi:10.1103/PhysRevD.99.043521
    [arXiv:1812.02333].

\bibitem{camarena18} 
  D.~Camarena and V.~Marra,
   ``Impact of the cosmic variance on $H_0$ on cosmological analyses,''
  Phys.\ Rev.\ D {\bf 98},  023537 (2018)
%   doi:10.1103/PhysRevD.98.023537
    [arXiv:1805.09900].

\bibitem{bengaly19a} 
  C.~A.~P.~Bengaly, U.~Andrade and J.~S.~Alcaniz,
  %``How does an incomplete sky coverage affect the Hubble Constant variance?,''
  Eur.\ Phys.\ J.\ C {\bf 79}, no. 9, 768 (2019)
%     doi:10.1140/epjc/s10052-019-7284-4
  [arXiv:1810.04966].

\bibitem{chen18} 
  H.~Y.~Chen, M.~Fishbach and D.~E.~Holz,
   ``A two per cent Hubble constant measurement from standard sirens within five years,''
  Nature {\bf 562}, 545 (2018)
%   doi:10.1038/s41586-018-0606-0
    [arXiv:1712.06531].
  
\bibitem{nair18} 
  R.~Nair, S.~Bose and T.~D.~Saini,
   ``Measuring the Hubble constant: Gravitational wave observations meet galaxy clustering,''
  Phys.\ Rev.\ D {\bf 98},  023502 (2018)
%   doi:10.1103/PhysRevD.98.023502
    [arXiv:1804.06085].
  
\bibitem{vitale18} 
  S.~Vitale and H.~Y.~Chen,
   ``Measuring the Hubble constant with neutron star black hole mergers,''
  Phys.\ Rev.\ Lett.\  {\bf 121}, 021303 (2018)
%   doi:10.1103/PhysRevLett.121.021303
    [arXiv:1804.07337].
  
\bibitem{mortlock18} 
  D.~J.~Mortlock, S.~M.~Feeney, H.~V.~Peiris, A.~R.~Williamson and S.~M.~Nissanke,
   ``Unbiased Hubble constant estimation from binary neutron star mergers,''
   Phys.\ Rev.\ D \textbf{100} (2019) no.10, 103523
  [arXiv:1811.11723].
  
\bibitem{shafieloo18} 
  A.~Shafieloo, R.~E.~Keeley and E.~V.~Linder,
   ``Will Gravitational Wave Sirens Determine the Hubble Constant?,''
   JCAP \textbf{03} (2020) no.03, 019
  [arXiv:1812.07775].
  
\bibitem{zhang19} 
  X.~Zhang,
  ``Gravitational wave standard sirens and cosmological parameter measurement,''
  Sci.\ China Phys.\ Mech.\ Astron.\  \textbf{62} (2019) no.11, 110431
  [arXiv:1905.11122].
    
\bibitem{feeney19} 
  S.~M.~Feeney, H.~V.~Peiris, A.~R.~Williamson, S.~M.~Nissanke, D.~J.~Mortlock, J.~Alsing and D.~Scolnic,
   ``Prospects for resolving the Hubble constant tension with standard sirens,''
  Phys.\ Rev.\ Lett.\  {\bf 122}, 061105 (2019)
%   doi:10.1103/PhysRevLett.122.061105
    [arXiv:1802.03404].

\bibitem{dominguez19} 
  A.~Dom\'inguez {\it et al.},
   ``A new measurement of the Hubble constant and matter content of the Universe using extragalactic background light $\gamma$-ray attenuation,''
    Astrophys.\ J.\  {\bf 881},  2 (2019)
  [arXiv:1903.12097].
  
\bibitem{fernandezarenas18} 
  D.~Fern\'andez Arenas {\it et al.},
  ``An independent determination of the local Hubble constant,''
  Mon.\ Not.\ Roy.\ Astron.\ Soc.\  {\bf 474},  1250 (2018)
%   doi:10.1093/mnras/stx2710
   [arXiv:1710.05951].
  
\bibitem{abbott18} 
  T.~M.~C.~Abbott {\it et al.} [DES Collaboration],
   ``Dark Energy Survey Year 1 Results: A Precise $H_0$ Measurement from DES Y1, BAO, and D/H Data,''
  Mon.\ Not.\ Roy.\ Astron.\ Soc.\  {\bf 480}, 3879 (2018)
%   doi:10.1093/mnras/sty1939
    [arXiv:1711.00403].

\bibitem{wang17a} 
  Y.~Wang, L.~Xu and G.~B.~Zhao,
  ``A measurement of the Hubble constant using galaxy redshift surveys,''
  Astrophys.\ J.\  {\bf 849},  84 (2017)
%   doi:10.3847/1538-4357/aa8f48
    [arXiv:1706.09149].  
  
 \bibitem{abbott17} 
  B.~P.~Abbott {\it et al.} [LIGO, Virgo, 1M2H, Dark Energy Camera GW-E, DES, DLT40, Las Cumbres, VINROUGE, MASTER Collaborations],
   ``A gravitational-wave standard siren measurement of the Hubble constant,''
  Nature {\bf 551},  85 (2017)
%   doi:10.1038/nature24471
    [arXiv:1710.05835].

\bibitem{keeley19} 
  R.~E.~Keeley, A.~Shafieloo, B.~L'Huillier and E.~V.~Linder,
  Mon. Not. Roy. Astron. Soc. \textbf{491} (2020) no.3, 3983-3989
%   doi:10.1093/mnras/stz3304
  [arXiv:1905.10216].

\bibitem{desi16} 
  A.~Aghamousa {\it et al.} [DESI Collaboration],
  ``The DESI Experiment Part I: Science,Targeting, and Survey Design,''
  arXiv:1611.00036.
  
\bibitem{Santos:2017qgq} 
  M.~G.~Santos {\it et al.} [MeerKLASS Collaboration],
  ``MeerKLASS: MeerKAT Large Area Synoptic Survey,''
  arXiv:1709.06099.
  
\bibitem{Santos2019}
M.~G.~Santos, private communication.
  
\end{thebibliography}
\end{document}